\documentclass{jpsj2}

\title{ Phase diagram of
the $p$-spin-interacting spin glass
with ferromagnetic bias
and a transverse field in the infinite-$p$ limit
}

\author{
Tomoyuki Obuchi$^1$, Hidetoshi Nishimori$^1$, and David Sherrington$^2$
}

\inst{
$^1$Department of Physics, Tokyo Institute of Technology,
Oh-okayama, Meguro-ku, Tokyo 152-8551\\
$^2$Rudolf Peierls Centre for Theoretical Physics, University of Oxford,\\
   1 Keble Road,
Oxford OX1 3NP, United Kingdom
}

\abst{
The phase diagram of the $p$-spin-interacting spin glass model in a
transverse field is investigated in the limit $p \rightarrow \infty$
under the presence of ferromagnetic bias.
Using the replica method and the static approximation, we show that
the phase diagram
consists of four phases:
   Quantum paramagnetic, classical paramagnetic, ferromagnetic, and
   spin-glass phases.
We also show that the static
approximation is valid in the ferromagnetic phase in the limit
$p \rightarrow \infty$ by using the large-$p$ expansion.
   Since the same approximation is already known to be valid
in other phases, we conclude that the obtained phase diagram is exact.
}

\kword{spin glass, $p$-spin interaction, random energy model, phase diagram, quantum effects}

\begin{document}
\maketitle

\section{Introduction}
A spin glass is a typical complex system characterized by quenched
disorder and frustration. Numerous studies of spin glass systems
clarified various interesting properties of disordered systems\cite{STAT,SPIN}.
Incorporation of quantum effects into disordered systems has been of
particular
interest and studied intensively\cite{Bray,Ishi,Butt,Thir,Gold}.
The noncommutativity of the operators
makes the problem interesting but difficult, and
a special care is required to obtain the correct result.
The Trotter decomposition
is known to be a useful approach to treat such effects\cite{Suzu}.
In this approach, order parameters become dependent on
the Trotter indices
and are determined self-consistently.

Bray and Moore\cite{Bray} proposed an approximate method to solve the problem.
Their method, which is referred to as the static approximation (SA), is
to neglect the time (or Trotter number) dependence of the order parameters.
Using the SA, Thirumalai et al.\cite{Thir} studied the
Sherrington-Kirkpatrick (SK) model \cite{SK}
in a transverse field to obtain the phase diagram.
They pointed out the limitation of the SA, showing that the entropy
does not vanish at zero temperature.

In order to understand the nature of quantum spin glasses,
it would be
helpful to investigate exactly solvable models.
The infinite-range spin glass model
with $p$-spin interactions is one of such tractable models.
In the limit $p \rightarrow \infty$,
this model is equivalent to
the so-called Derrida's random energy
model\cite{Derr}, which can be exactly solvable in a simple way
but retains nontrivial properties caused by quenched disorder.
Goldschmidt\cite{Gold}
investigated this model
in a transverse
field and
obtained the phase diagram, which consists of a spin glass (SG)
phase and
two paramagnetic phases: One is the classical paramagnetic (CP) phase
in which
quantum fluctuations are irrelevant and the other is the quantum
paramagnetic (QP) phase.
He also suggested that the SA is exact
in this model although there
is no rigorous proof.
Dobrosavljevic and Thirumalai\cite{Dobr} examined
the validity of the SA in the same model by performing a systematic large-$p$
expansion.
While they showed that the phase diagram of Goldschmidt is correct,
they also found that, for large but finite $p$, the SA is not valid in
the QP phase.

This model is an extreme simplification of SG models
but has a great
advantage to be exactly solvable,
which has enabled us to obtain many insights not only
about SG properties\cite{Gros}
but also about the replica method itself\cite{Derr}
and information processing problems\cite{Sour}.
Our main aim in this paper is to investigate what happens in this model in
a transverse field under the presence of ferromagnetic bias.
The influence of quantum fluctuations in the ferromagnetic (F) phase is
nontrivial, and it should be an interesting problem how the system
behaves as a result of three competing effects of disorder, quantum
fluctuations and ferromagnetic bias.
In \S\ref{sec2}, we calculate the free
energy of the model and give its phase diagram by using the
replica method and the SA.
In \S\ref{sec3},
we show the validity of the SA in the $p \rightarrow \infty$ limit
by using systematic large-$p$ expansion in the F phase.
The last section is devoted to conclusion.

\section{Replica analysis with the static approximation}\label{sec2}

\subsection{Replica symmetric free energy}
We consider the $p$-spin-interacting spin glass in a transverse
field. Evaluation of the partition function can be carried out by a
straightforward generalization of existing
methods\cite{Thir,Butt,Gold} to the case with
ferromagnetic bias.
The system is described
by the following Hamiltonian:
\begin{equation}
H=-\sum_{i_{1}<\ldots <i_{p}} J_{i_{1}\ldots i_{p}}
\sigma_{i_{1}}^{z} \ldots \sigma_{i_{p}}^{z} -\Gamma
\sum_{i=1}^{N} \sigma_{i}^{x} \equiv T+V,
\end{equation}
where $i$ is the site index, $\sigma^{z}$ and $\sigma^{x} $ are Pauli
spin operators, $\Gamma$ denotes the
strength of the transverse field and $ J_{i_{1}\ldots i_{p}}$ is the
quenched random interactions whose distribution function is given by
\begin{equation}
P(J_{i_{1}\ldots i_{p}}) = \left( \frac{N^{p-1}}{J^2 \pi p!}
\right)^{\frac{1}{2}}  \exp \left\{ -\frac{N^{p-1}}{J^2 p!}
\left(J_{i_{1}\ldots i_{p}} - \frac{ j_{0} p! }{ N^{p-1} } \right)^2 \right\}.
\end{equation}
The limit $p \rightarrow  \infty$ is taken after all calculations.
The partition function of this quantum system can be represented in terms
   of a corresponding classical spin system with the Ising variable
   $\sigma= (\pm 1)$
by the Trotter decomposition\cite{Suzu}
\begin{equation}
Z=\lim_{M \rightarrow \infty} {\rm Tr} \left({\rm e}^{-\beta T/M} {\rm e}^{-\beta V/M}
\right)^M=\lim_{M \rightarrow \infty} Z_{M},
\end{equation}
where
\begin{equation}
Z_{M}=C^{MN}{\rm Tr} \exp \left( \frac{
\beta }{M}\sum_{t=1}^{M}\sum_{i_{1}<\ldots <i_{p}}J_{i_{1}\ldots
i_{p}} \sigma_{i_{1},t} \ldots \sigma_{i_{p},t} + B
\sum_{t=1}^{M}\sum_{i}\sigma_{i,t}\sigma_{i,t+1} \right),
\end{equation}
and the constants $B$ and $C$ are related to the transverse field
$\Gamma$ as
\begin{equation}
   B = \frac{1}{2}\ln \coth\frac{\beta \Gamma}{M}, \ \ C=\left(
\frac{1}{2}\sinh\frac{2\beta \Gamma}{M} \right)^{\frac{1}{2}}.
\end{equation}
The symbol ${\rm Tr}$ denotes the trace over the $\sigma$-spins.

We use the replica method \cite{SK}:
\begin{equation}
[\log Z]=\lim_{n \rightarrow 0} \frac{[Z^n]-1}{n}
\end{equation}
to carry out the random interactions
$J_{i_{1}\ldots i_{p}}$ averages $[\cdots]$.
The replicated partition function is given by
\begin{eqnarray}
\left[ Z^{n}_{M} \right] =
   {\rm Tr} \exp
   \Biggl( \frac{\beta^2 J^2 N}{4M^2}\sum_{t,t'=1}^{M} \sum_{\mu, \nu =1}^{n}
    \left(\frac{1}{N}\sum_{i}\sigma_{i,t}^{\mu} \sigma_{i,t'}^{\nu} \right)^p
    +  \frac{\beta j_{0} N}{M}\sum_{t=1}^{M}\sum_{\mu=1}^{n}
    \left(\frac{1}{N}\sum_{i}\sigma_{i,t}^{\mu} \right)^{p} \notag \\
    + B\sum_{t=1}^{M} \sum_{\mu=1}^{n}\sum_{i}\sigma_{i,t}^{\mu}\sigma_{i,t+1}^{\mu}
   \Biggr),
\end{eqnarray}
where the replica indices are denoted by $\mu$ and $\nu$.
We have omitted some irrelevant constants.
The spin product term $\left(\sum_{i}\sigma_{i,t}^{\mu}
\sigma_{i,t'}^{\nu}/N \right)^p$
can be simplified by introducing an order parameter
$q_{tt'}^{\mu\nu}$ and its conjugate Lagrange multiplier
${\widetilde q_{tt'}^{\mu\nu} }$ for the constraint
$q_{tt'}^{\mu\nu} =\sum_{i}\sigma_{i,t}^{\mu} \sigma_{i,t'}^{\nu}/N $.
In the present case,  we must distinguish
diagonal $q^{\mu\mu}_{tt'}$ and off-diagonal
$q^{\mu\nu}_{tt'} (\mu \neq \nu)$. Physically,
$q^{\mu\mu}_{tt'}$ is a measure of quantum fluctuations and
$q_{tt'}^{\mu\nu}$ is the SG order parameter. If there are no quantum
fluctuations, the spin configuration $\sigma_{i,t}^{\mu}$ does not depend
on time $t$ and $q^{\mu\mu}_{tt'}=1$. Quantum fluctuations
reduce $q^{\mu\mu}_{tt'}$ from unity. Hence, $1-q^{\mu\mu}_{tt'}$ gives
a measure of quantum fluctuations.
Also, to simplify $\left(\sum_{i}\sigma_{i,t}^{\mu}/N\right)^p$,
the ferromagnetic order parameter $m_{t}^{\mu}$ is introduced and the
constraint $m_{t}^{\mu}=\sum_{i}\sigma_{i,t}^{\mu}/N$ is imposed by the
integration over the conjugate variable ${\widetilde m}_{t}^{\mu}$.
Using these notations,
we can rewrite the effective partition function as
\begin{eqnarray}
   &&\left[ Z^{n}_{M} \right] =
   \int \prod_{\mu}\prod_{t}      dm_{t}^{\mu}       d{\widetilde m_{t}^{\mu}}
   \prod_{\mu < \nu}\prod_{t,t'}  dq_{tt'}^{\mu\nu}  d{\widetilde
q_{tt'}^{\mu\nu}}
   \prod_{\mu}\prod_{t \neq t'}   dq_{tt'}^{\mu\mu}  d{\widetilde
q_{tt'}^{\mu\mu} }  \notag \\
   && \times \exp N \Biggl\{
    \sum_{t,t'} \sum_{\mu < \nu } \left( \frac{\beta^2 J^2 }{2
M^2}\left(  q_{tt'}^{\mu\nu} \right)^p -
				 \frac{1}{M^2} {\widetilde q_{tt'}^{\mu\nu}}q_{tt'}^{\mu\nu}  \right)
    + \sum_{t,t'} \sum_{\mu } \left( \frac{\beta^2 J^2 }{4 M^2} \left(
q_{tt'}^{\mu\mu} \right)^p -
			     \frac{1}{M^2} { \widetilde q_{tt'}^{\mu\mu}}q_{tt'}^{\mu\mu}   \right) \notag \\
   & & + \sum_{t} \sum_{\mu } \left( \frac{\beta^2 j_{0} }{M} \left(
m_{t}^{\mu} \right)^p -
			     \frac{1}{M}{ \widetilde m_{t}^{\mu}}m_{t}^{\mu}   \right)
   + \log {\rm Tr} \exp \left( -H_{{\rm eff}} \right)
   \Biggr\} ,
\end{eqnarray}
where
\begin{equation}
H_{{\rm eff}}=
   - B \sum_{t} \sum_{\mu}  \sigma_{t}^{ \mu }\sigma_{ t+1 }^{ \mu }
     - \frac{1}{M}   \sum_{\mu} \sum_{t} { \widetilde m^{\mu}_{t} }
\sigma_{t}^{\mu}
     - \frac{1}{M^2} \sum_{\mu < \nu}\sum_{t,t'} { \widetilde
q^{\mu\nu}_{tt'} } \sigma_{t}^{\mu}\sigma_{t'}^{\nu}
     - \frac{1}{M^2} \sum_{\mu} \sum_{t \neq t'} { \widetilde
q^{\mu\mu}_{tt'} } \sigma_{t}^{\mu}\sigma_{t'}^{\mu}.
\label{eq:Heff}
\end{equation}
We can calculate the free energy per spin $F$ of the replicated systems
in the thermodynamic limit by the saddle-point
method. The result is
\begin{eqnarray}
   -\beta F=
    \sum_{t,t'} \sum_{\mu < \nu }
    \left( \frac{\beta^2 J^2 }{2 M^2}\left(  q_{tt'}^{\mu\nu} \right)^p -
     \frac{1}{M^2} {\widetilde q_{tt'}^{\mu\nu}}q_{tt'}^{\mu\nu}  \right)
    + \sum_{t,t'} \sum_{\mu }
    \left( \frac{\beta^2 J^2 }{4 M^2}\left( q_{tt'}^{\mu\mu} \right)^p -
     \frac{1}{M^2} { \widetilde q_{tt'}^{\mu\mu}}q_{tt'}^{\mu\mu}\right) \notag \\
   +\sum_{t} \sum_{\mu }
    \left( \frac{\beta j_{0} }{M} \left( m_{t}^{\mu} \right)^p -
     \frac{1}{M}{ \widetilde m_{t}^{\mu}}m_{t}^{\mu}   \right)
    +\log {\rm Tr} \exp \left( -H_{{\rm eff}} \right),
\label{eq:rep fe}
\end{eqnarray}
where
\begin{subequations}
\begin{align}
q_{tt'}^{\mu \nu}=\langle \sigma_{t}^{\mu}\sigma_{t'}^{\nu} \rangle \ , \
{\widetilde q_{tt'}^{\mu\nu} } = \frac{1}{2}\beta^2 J^2 p(
q_{tt'}^{\mu \nu} )^{p-1},\label{eq:state primary cq} \\
q_{tt'}^{\mu \mu}=\langle \sigma_{t}^{\mu}\sigma_{t'}^{\mu} \rangle \ , \
{\widetilde q_{tt'}^{\mu\mu} } = \frac{1}{4}\beta^2 J^2 p(
q_{tt'}^{\mu \mu} )^{p-1},  \\
m_{t}^{\mu}=\langle \sigma_{t}^{\mu} \rangle \ , \
{\widetilde m_{t}^{\mu}}=\beta j_{0}
p(m_{t}^{\mu})^{p-1}. \label{eq:state primary cm}
\end{align}
\end{subequations}
The brackets $\langle \cdots \rangle $ denote the average by the
weight $\exp(-H_{{\rm eff}})$.
It is difficult to solve these equations exactly for arbitrary values of
$p$ because of the time dependence of the
order parameters. However, in the limit
$p \rightarrow \infty$, the problem is considerably simplified because
conjugate variables can be either 0 or $\infty $. For example,
if we restrict ourselves to the case that $m_{t}^{\mu}$ is non-negative
(the other case can be treated similarly),
eq. (\ref{eq:state primary cm}) implies $0 \leq m_{t}^{\mu} \leq 1$, which
leads to  either
${\widetilde m}_{t}^{\mu} \rightarrow 0 \,\, (0 \leq m_{t}^{\mu} < 1)$
or
${\widetilde m}_{t}^{\mu} \rightarrow \pm \infty \,\, (m_{t}^{\mu}
\rightarrow 1)$.
Therefore, the SA gives
the exact phase boundaries as is shown below.


Now, we assume the replica symmetry (RS) and use the SA:
\begin{equation}
   q^{\mu\mu}_{tt'} \equiv R \ ,\  q^{\mu\nu}_{tt'} \equiv q \ ,\
    m^{\mu}_{t} \equiv m.
\end{equation}
As already noted, $R$ is the order parameter measuring the effect of
quantum fluctuations, $q$ is the conventional SG order parameter
and $m$ denotes the ferromagnetic order parameter. The free energy
is then reduced to the expression
\begin{eqnarray}
&&\!\!\!\!\!\!\!\!\!\!\!\!\!\!\!\!\!\!\!\!\!\!\!\!-\beta f \equiv
-\lim_{M \rightarrow \infty}\lim_{n \rightarrow 0} \frac{\beta F}{n}  \notag \\
&&\!\!\!\!\!\!\!\!\!\! =\frac{1}{4}\beta^2 J^2\left(R^p -q^p \right) +
   \frac{1}{2}{\widetilde q}q-{\widetilde R}R
   +\beta j_{0} m^p -{\widetilde m}m +\lim_{M \rightarrow
\infty}\lim_{n \rightarrow 0}\frac{1}{n} \log {\rm Tr} \exp\left(-H_{{\rm
eff}}\right).
\end{eqnarray}
Since all order parameters and the conjugate variables are
independent of time and replica indices,
   the summation of spin products
in $H_{ {\rm eff} }$ is rewritten as
\begin{equation}
\sum_{\mu \neq
\nu}\sum_{t,t'}\sigma^{\mu}_{t}\sigma^{\nu}_{t'}=\frac{1}{2}\left\{
\left(\sum_{\mu}\sum_{t} \sigma_{t}^{\mu} \right)^2 -
\sum_{\mu}\left(\sum_{t}\sigma_{t}^{\mu}\right)^2 \right\}.
\end{equation}
Then, the effective Boltzmann
factor $\exp(- H_{ {\rm eff}} )$
is considerably simplified by the Hubbard-Stratonovich
transformation. We obtain
\begin{equation}
\exp(-H_{{\rm eff}})=\int Dz_{1}\prod_{\mu}\left\{ \int Dz_{2}
\exp\left(
B\sum_{t}\sigma_{t}^{\mu}\sigma_{t+1}^{\mu}+\frac{A}{M}\sum_{t}\sigma_{t}^{\mu}
\right) \right\}, \label{eq:Heff}
\end{equation}
where
\begin{equation}
A=\sqrt{2{\widetilde R}-{\widetilde q}} \, z_{2}
+\sqrt{{\widetilde q}} \, z_{1}+{\widetilde m}\ , \
Dz=\frac{dz}{\sqrt{2\pi}}{\rm e}^{-\frac{1}{2}z^2}.\label{eq:A}
\end{equation}
Using the Trotter formula,
we can take the spin trace in
the limit $M \rightarrow \infty$ as
\begin{equation}
\lim_{M \rightarrow \infty }
{\rm Tr} \exp\left(
B\sum_{t}\sigma_{t}\sigma_{t+1}+\frac{A}{M}\sum_{t}\sigma_{t} \right)
={\rm Tr} {\rm e}^{A\sigma^{z}+\beta \Gamma \sigma^{x}} =2\cosh\sqrt{A^2+\beta^2
\Gamma^2}.\label{eq:Trotter}
\end{equation}
Because each replica gives the same contribution to the replicated free
energy,  the limit of $n \rightarrow 0$ is easily taken.
The result is
\begin{eqnarray}
-\beta f =
   \frac{1}{4}\beta^2 J^2\left(R^p -q^p \right) +
   \frac{1}{2}{\widetilde q}q-{\widetilde R}R
   +\beta j_{0} m^p -{\widetilde m}m +\int Dz_{1} \,\log \int Dz_{2}\,
(2\cosh \omega), \label{eq:fesa}
\end{eqnarray}
where
\begin{equation}
   \omega = \left( A^2 + \beta ^2 \Gamma ^2\right)^{\frac{1}{2}}. \label{eq:omega}
\end{equation}
The saddle-point conditions of the free energy are
\begin{eqnarray} 
& & {\widetilde m}=\beta j_{0}pm^{p-1}, \label{eq:state cm}\\
& & {\widetilde R}=\frac{1}{4}\beta^2 J^2 p R^{p-1},
\label{eq:state cc}\ \\
& & {\widetilde q}=\frac{1}{2}\beta^2 J^2 p q^{p-1},  \label{eq:state cq} \\
& & m=\int Dz_{1} \, Y^{-1}\int Dz_{2} \, A\omega^{-1} \sinh \omega,
\label{eq:state m}\\\
& & R=\int Dz_{1} \, Y^{-1}
   \left( \int Dz_{2} \, A^2\omega^{-2}\cosh\omega +
    \beta^2 \Gamma^2 \int Dz_{2} \, \omega^{-3}\sinh \omega \right),
\label{eq:state c}\\
& & q=\int Dz_{1} \, Y^{-2} \left( \int Dz_{2} \, A\omega^{-1} \sinh
\omega \right)^2,  \label{eq:state q}
\end{eqnarray}
where
\begin{equation}
Y=\int Dz_{2} \, \cosh \omega. \label{eq:Y}
\end{equation}
When $\Gamma$ is equal to 0, $R$ becomes unity and the free energy
is reduced to the classical result\cite{Derr,Gros}.
These eqs. (\ref{eq:state cm})-(\ref{eq:state q}) generalize the
result of Goldschmidt\cite{Gold} to the case of finite $j_{0}$.

\subsection{Solutions of the equations of state.}
The following inequality is useful to evaluate
the solutions of
eqs. (\ref{eq:state cm})-(\ref{eq:state q}).\cite{Nish}
\begin{eqnarray}
   R & = & \int Dz_{1} \, Y^{-1}
   \left( \int Dz_{2} \, A^2\omega^{-2}\cosh\omega +
    \beta^2 \Gamma^2 \int Dz_{2} \, \omega^{-3}\sinh \omega \right) \notag \\
& & \geq \int Dz_{1} \, Y^{-1} \int Dz_{2} \, A^2\omega^{-2}\cosh\omega
    \geq \int Dz_{1} \, Y^{-1}\int Dz_{2} \, A^2\omega^{-2}\sinh\omega  \notag \\
& & \geq \int Dz_{1} \, Y^{-2} \left( \int Dz_{2} \, A\omega^{-1}
\sinh \omega \right)^2=q \label{eq:c dainari q}.
\end{eqnarray}
  From eqs. (\ref{eq:state cm})-(\ref{eq:state cq}) and $0 \le R,q,m \le 1$,
conjugate variables can be either 0 or $\infty$ in the limit
$p \rightarrow \infty$. These conditions restrict combinations of
the values of conjugate variables to
\begin{equation}
({\widetilde m},{\widetilde R},{\widetilde
q})=(0,0,0),(0,\infty,0),(0,\infty,\infty),(\infty,0,0),(\infty,\infty,0),(\infty,\infty,\infty).
\end{equation}
We must exclude the solution
$({\widetilde m},{\widetilde R},{\widetilde q})=(\infty,0,0)$ as follows.
Substituting these values $(\infty,0,0)$ into eqs. (\ref{eq:state 
m})-(\ref{eq:state q}),
we find the solution $(m,R,q)=(1,1,1)$, which
contradicts the condition $({\widetilde R},{\widetilde q})=(0,0)$
as one can check from eqs. (\ref{eq:state cc}) and (\ref{eq:state 
cq}) with $p\to\infty$.
Similarly, $({\widetilde m},{\widetilde R},{\widetilde q})=(\infty,\infty,0)$ is
also inappropriate. Finally, we get four solutions
\begin{equation}
(m,R,q)=(0,\frac{1}{\beta \Gamma}\tanh\beta\Gamma ,0),(0,1,0),(0,1,1),(1,1,1).
\end{equation}

The solution $(m,R,q)=(0,1,0)$ is a paramagnetic one which is
identical to that in the case without quantum fluctuations. In this
solution quantum
effects are irrelevant so that this phase is the CP phase.
The solution $(m,R,q)=(0,\frac{1}{\beta \Gamma}\tanh\beta\Gamma,0)$
represents a non-trivial paramagnetic phase
which arises due to quantum fluctuations. $R$ is
reduced from unity by the transverse field $\Gamma$ and
this phase is the QP phase.
The solution $(m,R,q)=(0,1,1)$ is for the SG phase. In the limit
   $p \rightarrow \infty$, the finite value solution of $q$ is limited to
$q=1$ from eqs. (\ref{eq:state cm})-(\ref{eq:state cq}). From the inequality
   $R \ge q$, $R$ is also equal to $1$
and hence we expect that quantum fluctuations are
irrelevant in this phase.\cite{Dobr,Cesa2}
The solution $(m,R,q)=(1,1,1)$ is ferromagnetic.
   In this phase, all
   order parameters are restricted to $1$ in the limit
$p \rightarrow \infty$. Therefore, quantum
fluctuations are also irrelevant
in this phase as will also be shown below. We summarize the above
results in Table \ref{tab:order}.
\begin{table}[hbt]
    \begin{center}
\begin{tabular}{c|c|c}
\noalign{\hrule height0.8pt}
Phase & ($m,R,q$) & (${\widetilde m},{\widetilde R},{\widetilde q}$) \\
\hline
CP & ($0,1,0$) & ($0,\infty,0$) \\
QP & ($0,\tanh \beta \Gamma/(\beta \Gamma),0$) & ($0,0,0$) \\
F & ($1,1,1$) & ($\infty,\infty,\infty$) \\
SG & ($0,1,1$) & ($0,\infty,\infty$) \\
\noalign{\hrule height0.8pt}
\end{tabular}
\end{center}
\caption{Values of order parameters in various phases.} \label{tab:order}
\end{table}

Substituting the values of the order parameters and
conjugate variables into eq. (\ref{eq:fesa}), we obtain
the corresponding free energy as
\begin{eqnarray}
&&f_{{\rm QP}}=-T\log2 -T\log\cosh \beta \Gamma \label{eq:feqp}, \\
&&f_{{\rm CP}}=-\frac{1}{4}\beta J^2 -T\log2 \label{eq:fecp}, \\
&&f_{{\rm F}}=-j_{0} \label{eq:fef}, \\
&&f_{{\rm SG}}\rightarrow -T\sqrt{ \frac{ 2{\widetilde q} }{\pi} }
\rightarrow -\infty \ \ \ \ \ (p \rightarrow \infty). \label{eq:fesg}
\end{eqnarray}
For this large $p$ case the free energy of the SG phase is always smaller than
the other free energies in eqs. (\ref{eq:feqp}), (\ref{eq:fecp}) and 
(\ref{eq:fef}), although
one knows that in spin glass phases maximization is the correct 
procedure\cite{STAT}.
However, this feature is an artifact of the RS solution.

\subsection{Free energy in the spin glass phase}
The correct solution of the SG phase is derived by the RSB.
In the limit $p \rightarrow \infty$, the first step of the RSB (1RSB)
is known to be sufficient in the classical case ($\Gamma =0$)\cite{Gros}.
We may expect that the same is true in the presence of a transverse field
and therefore discuss the 1RSB scheme here.
Note that at sufficiently low temperatures the full-step RSB is known 
to be necessary for finite $p$
\cite{Gardner,GNS}, which we do not take into account explicitly here
because we take the limit $p\to\infty$ in the end.
New order parameters
and a branch-point parameter $m_{1}$ are defined as follows
\begin{equation}
q^{l \mu_{l},l \nu_{l}}=q_{0} \ , \ q^{l \mu_{l},l' \nu_{l'}}=q_{1}
\ \ \ \ \ l\neq l',
\end{equation}
where $l=1,\ldots,n/m_{1} $ is the block number and
$\mu_{l}=1,\ldots,m_{1}$ is
the index inside a block.
Detailed calculations are given in
Appendix A. The 1RSB free energy is found to be given as
\begin{eqnarray}
&&\beta f=\frac{1}{2}m_{1}\left(\frac{1}{2}\beta^2 J^2 (q_{0})^p
			   - {\widetilde q_{0}}q_{0}\right) +
\frac{1}{2}(1-m_{1})\left(\frac{1}{2}\beta^2 J^2 (q_{1})^p
		     - {\widetilde q_{1}}q_{1}\right) \notag \\
&& -\frac{1}{4}\beta^2 J^2 (R)^{p}+ {\widetilde R}R - \beta
j_{0}m^{p}+{\widetilde m}m
   -\frac{1}{m_{1}}\int Dz_{1} \, \log \int Dz_{2} \left(\int Dz_{3}
\, 2\cosh \omega_{1} \right)^{m_{1}}\!\!\! ,\label{eq:fe1rsb}
\end{eqnarray}
where
\begin{equation}
\omega_{1} = \left( A_{1}^2 + \beta ^2 \Gamma ^2 \right)^{\frac{1}{2}}\ ,\
    A_{1}= \sqrt{ { \widetilde q_{0} } } \, z_{1}
    +\sqrt{ {\widetilde q_{1}}- { \widetilde q_{0} } }\, z_{2}
    +\sqrt{ 2{\widetilde R}-{\widetilde q_{1} } } \, z_{3} \,\,
    +{\widetilde m}.\label{eq:A1}
\end{equation}
If we set $q_{0}=q_{1}$, the RS solution (\ref{eq:fesa}) is reproduced.
The saddle-point conditions of the 1RSB free energy are given by
\begin{eqnarray}
& &\!\!\!\!\!\!\!\! {\widetilde m}=\beta j_{0}pm^{p-1}, \label{eq:state cmrsb}\\
& &\!\!\!\!\!\!\!\! {\widetilde R}=\frac{1}{4}\beta^2 J^2 p R^{p-1}, \\
& &\!\!\!\!\!\!\!\! {\widetilde q}_{0}=\frac{1}{2}\beta^2 J^2 p q^{p-1}_{0},  \\
& &\!\!\!\!\!\!\!\! {\widetilde q}_{1}=\frac{1}{2}\beta^2 J^2 p q^{p-1}_{1},  \\
& &\!\!\!\!\!\!\!\! m=\int Dz_{1} \, Y^{-1}_{1}\int Dz_{2} \,Y_{2}^{m_{1}-1},
   \int Dz_{3} \, A_{1}\omega^{-1}_{1} \sinh \omega_{1}, \label{eq:state mrsb}\\
& &\!\!\!\!\!\!\!\! R=\int Dz_{1} \, Y^{-1}_{1}
   \int Dz_{2} \, Y_{2}^{m_{1}-1}\left(\int Dz_{3} \,
		   A^2_{1}\omega^{-2}_{1} \cosh\omega_{1} +
    \beta^2 \Gamma^2 \int Dz_{3} \, \omega^{-3}_{1}\sinh \omega_{1}
\right), \label{eq:state crsb}\\
& &\!\!\!\!\!\!\!\! q_{0}=\int Dz_{1} \, \left( Y^{-1}_{1} \int
Dz_{2} \, Y_{2}^{m_{1}-1}
		       \int Dz_{3} \,
		       A_{1}\omega^{-1}_{1}\sinh \omega_{1}\right)^2, \label{eq:state q0rsb}\\
& &\!\!\!\!\!\!\!\! q_{1}=\int Dz_{1} \,Y^{-1}_{1} \int Dz_{2} \, Y_{2}^{m_{1}-2}
   \left( \int Dz_{3} \, A_{1}\omega^{-1}_{1} \sinh \omega_{1}
\right)^2, \label{eq:state q1rsb}
\end{eqnarray}
where
\begin{equation}
Y_{1}=Y_{1}(z_{1})\equiv \int Dz_{2}\left( \int Dz_{3} \,
   \cosh \omega_{1} \right)^{m_{1}}
\ ,\ Y_{2}=Y_{2}(z_{1},z_{2})\equiv \int Dz_{3} \,\cosh \omega_{1}.
\label{eq:Y1}
\end{equation}
   Inequalities $R \geq q_{1} \geq q_{0}$ are also derived in a similar
way to the derivation of eq. (\ref{eq:c dainari q})
as explained in detail in Appendix A. In the limit
$p \rightarrow \infty$, we find from these inequalities and
eqs. (\ref{eq:state cmrsb})-(\ref{eq:state q1rsb}) that the only
non-trivial RSB solution is
$(m,R,q_{0},q_{1})=(0,1,0,1)$.
The corresponding free energy is
\begin{equation}
f=-\frac{1}{4}\beta J^2 m_{1} - T\frac{1}{m_{1}}\log2.
\end{equation}
Taking a variation with respect to $m_{1}$, we find
\begin{equation}
f_{{\rm SG}}=-J\sqrt{\log2}.
\end{equation}
This is the correct free energy of the SG phase.

\begin{figure}[htbp]
\begin{minipage}{0.5\hsize}
\begin{center}
     \includegraphics[width=60mm]{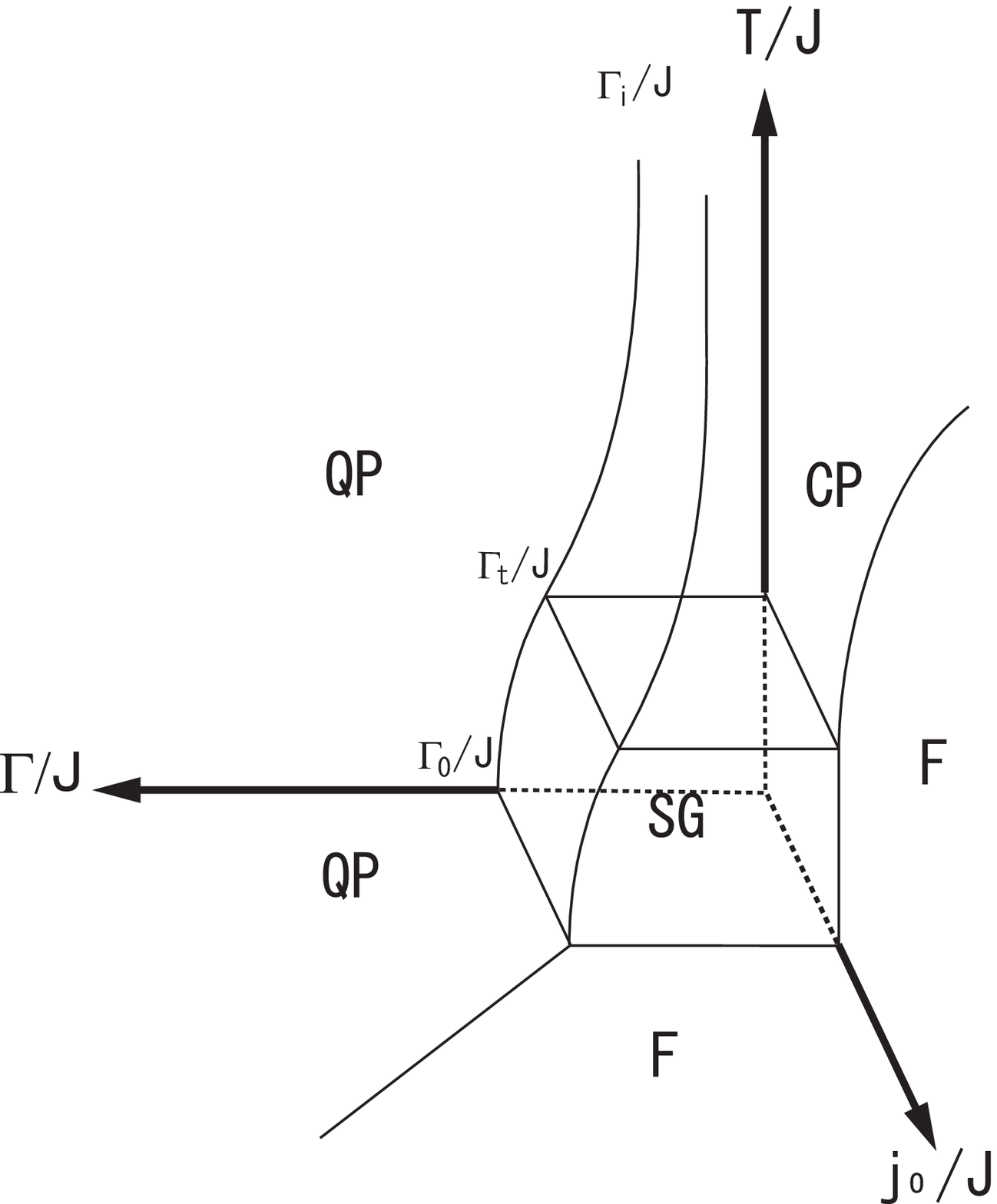}
   \caption{Full phase diagram of the model
in the limit $p \rightarrow \infty$. The QP phase
   appears when $\Gamma/J$ becomes larger than
   $\Gamma_{i}/J=1/\sqrt{2}$. The CP phase is completely suppressed by
   quantum fluctuations in the range of
   $\Gamma/J \geq \Gamma_{t}/J=\log(2+\sqrt{3})/(2\sqrt{2})$.
   Replica-symmetry breaking exists only in the SG phase.}
   \label{fig:p-d}
\end{center}
\end{minipage}
\hspace{3mm}
   \begin{minipage}{0.5\hsize}
    \begin{center}
     \includegraphics[width=70mm]{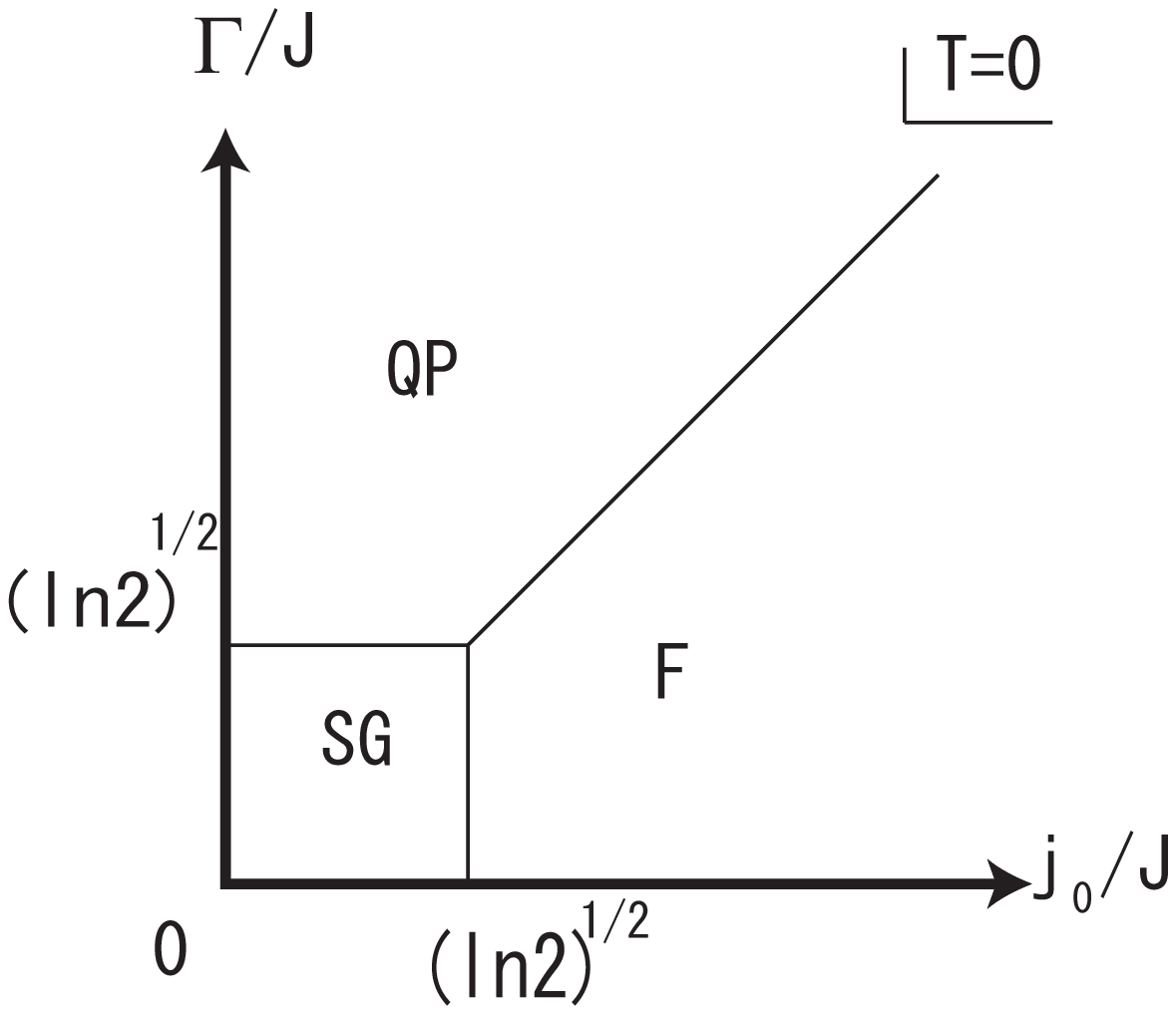}
     \caption{Ground-state phase diagram on the $T=0$ plane of
     Fig. \ref{fig:p-d}.  For large $\Gamma$,
     quantum fluctuations destroy the ferromagnetic order and
     cause a phase transition to the QP phase.}
     \label{fig:g-p-d}
    \end{center}
   \end{minipage}
\end{figure}

\begin{figure}[htbp]
\begin{minipage}{0.5\hsize}
\begin{center}
     \includegraphics[width=60mm]{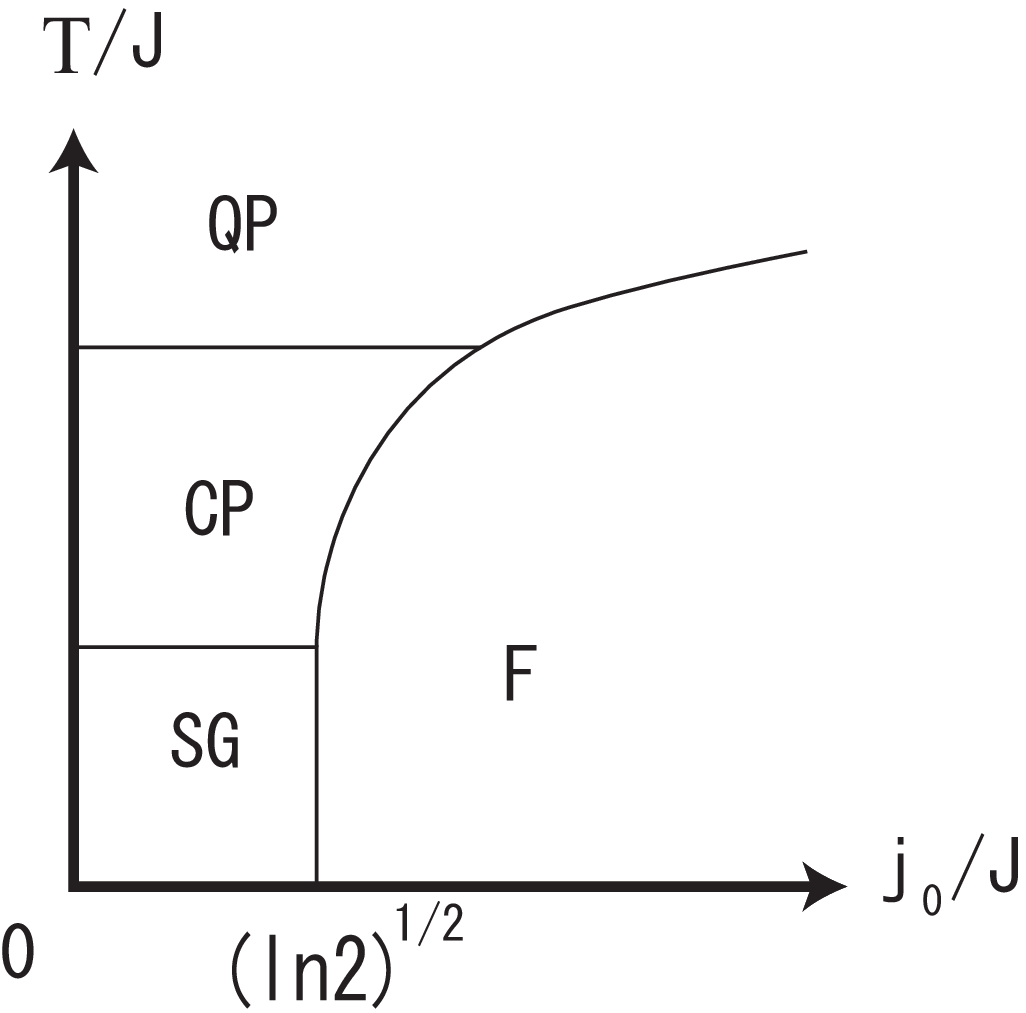}
   \caption{Schematic phase diagram on the $T$-$j_{0}$ plane in the range of
   $\Gamma_{i} \leq \Gamma \leq \Gamma_{t}$. For large $T$, the QP phase
   appears instead of the CP phase. As $\Gamma$ grows, the CP phase
   diminishes and vanishes at $\Gamma =\Gamma_{t}$. }
   \label{fig:T-j-p-d}
\end{center}
\end{minipage}
\hspace{3mm}
   \begin{minipage}{0.5\hsize}
    \begin{center}
     \includegraphics[width=50mm]{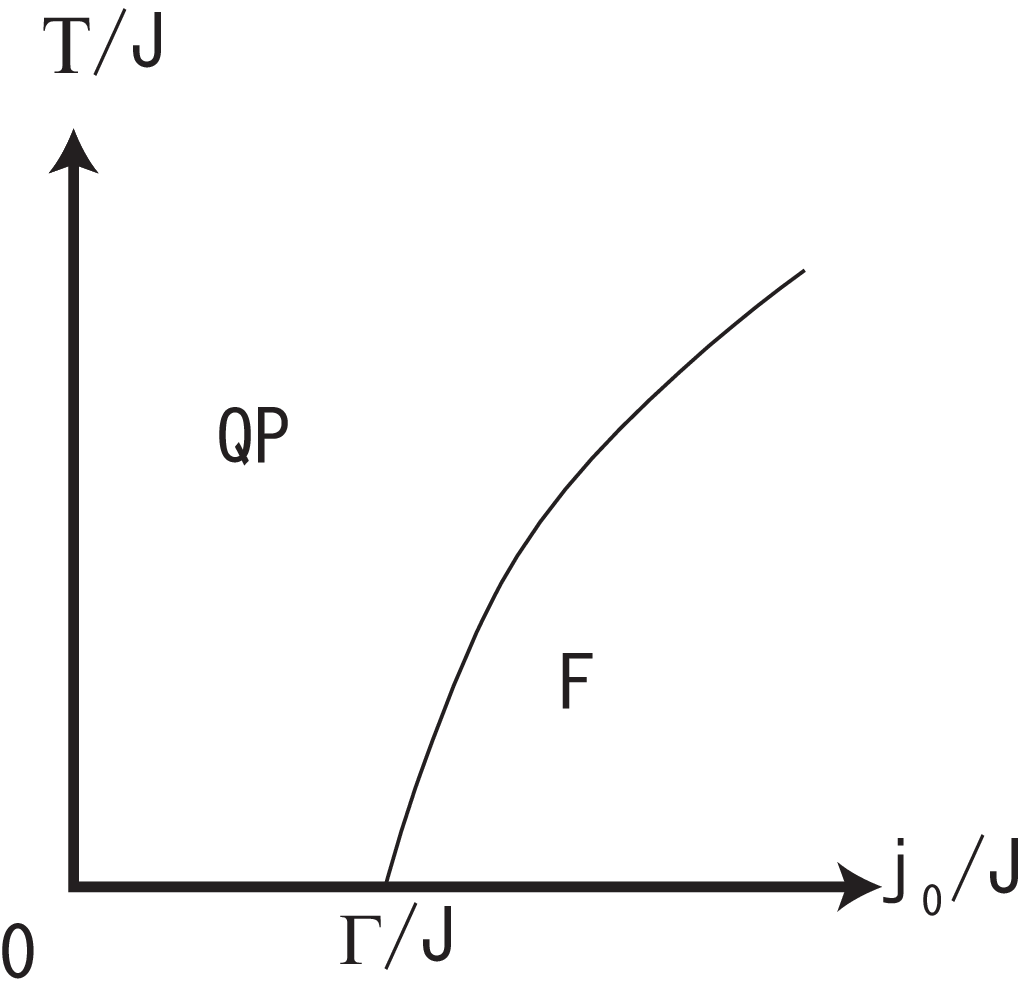}
     \caption{Phase diagram on the $T$-$j_{0}$ plane for large $\Gamma$.
     The SG phase disappears at $\Gamma/J=\Gamma_{0}/j=\sqrt{\log2}$. }
     \label{fig:T-j2-p-d}
    \end{center}
   \end{minipage}
\end{figure}

We can determine all the
phase boundaries by equating free energies between different phases. The
phase diagram thus obtained is shown in Figs. \ref{fig:p-d}-\ref{fig:T-j2-p-d}.
We can see that, as $\Gamma$ grows,
quantum fluctuations reduce the ferromagnetic order and
cause a phase transition to the QP phase as in Fig. \ref{fig:g-p-d}.
Order parameters discontinuously change at any phase boundary.
In that sense, all the phase transitions are of first order.

\section{Validity of the static ansatz}\label{sec3}

\subsection{Expansion from the large-$p$ limit}
In this section, we check the validity of the SA. Again, the method is
generalization of ref. $11$ to the case with ferromagnetic bias. For
that purpose, we introduce
corrections to the SA with $t,t'$-dependence and expand the free energy
with respect to those correction terms.
Then, it is shown that
the time-dependent parts are irrelevant
   in the limit $p \rightarrow \infty$.

We start from the RS solution
\begin{equation}
q_{t,t'}^{\mu\mu}=R(t,t') \ ,\ q_{t,t'}^{\mu \nu}=q(t,t')\ ,
\ m^{\mu}_{t}=m(t).
\end{equation}
Separating
each conjugate variable to the static
and time-dependent parts, we rewrite the effective Hamiltonian
$H_{{\rm eff}}=H_{{\rm stat}}+V(t,t')$ as
\begin{eqnarray}
H_{{\rm stat}}=
    -B\sum_{\mu}\sum_{t} \sigma_{t}^{\mu}\sigma_{t+1}^{\mu}
    -\frac{1}{M^2}{\widetilde R}\sum_{\mu}\sum_{t \neq t'}
\sigma_{t}^{\mu}\sigma_{t'}^{\mu}
    -\frac{1}{2M^2}{\widetilde q}\sum_{\mu \neq \nu}\sum_{t,t'}
\sigma_{t}^{\mu}\sigma_{t'}^{\nu}
   -\frac{1}{M}{\widetilde m}\sum_{\mu}\sum_{t}      \sigma_{t}^{\mu}, \\
   V(t,t')=-\frac{1}{M^2}\sum_{\mu}\sum_{t \neq t'}
\Delta{\widetilde R}(t,t') \sigma_{t}^{\mu}\sigma_{t'}^{\mu}
    -\frac{1}{2M^2}\sum_{\mu \neq \nu}\sum_{t,t'}\Delta{\widetilde
q}(t,t')\sigma_{t}^{\mu}\sigma_{t'}^{\nu}
    -\frac{1}{M}\sum_{\mu}\sum_{t}\Delta{\widetilde m}(t)\sigma_{t}^{\mu},
   \label{eq:Heff(t)}
\end{eqnarray}
where
\begin{equation}
{\widetilde R}(t,t')={\widetilde R}+\Delta{\widetilde R}(t,t')\ ,\
{\widetilde q}(t,t')={\widetilde q}+\Delta{\widetilde q}(t,t')\ ,\
{\widetilde m}(t,t')={\widetilde m}+\Delta{\widetilde m}(t,t').
\end{equation}
It is expected that the order parameters are monotone decreasing
functions of the time interval $|t-t'|$ because they are
originally
   written as spin-correlation functions
eqs. (\ref{eq:state primary cq})-(\ref{eq:state primary cm}).
Then, the time-dependent parts of their
conjugate variables, which are the $p$th powers
of the order parameters, become drastically small for large $p$.
Therefore,
it is reasonable to expand the free energy with respect to
the time-dependent part.
The free energy is expanded with respect to $V(t,t')$ to first order as
\begin{eqnarray}
   &&\!\!\!\!\!\!\!\!\!\!\!\!\!\!\!\!\!\!\!\!\!\! \beta
f=\frac{1}{2}\frac{1}{M^2}\sum_{t,t'}
    \left(\frac{1}{2}\beta^2 J^2 q(t,t')^p- {\widetilde
q}q(t,t')-\Delta {\widetilde q}(t,t')q(t,t')\right) \notag \\
   &&\!\!\!\!\!\!\!\! -\frac{1}{M^2}\sum_{t \neq t'}
    \left(\frac{1}{4}\beta^2 J^2 R(t,t')^p- {\widetilde
R}R(t,t')-\Delta {\widetilde R}(t,t')R(t,t')\right) \notag \\
   &&\!\!\!\!\!\!\!\!\!\!\!\!\! -\frac{1}{M}\sum_{t}
    \left(\beta j_{0} m(t)^p- {\widetilde m}m(t)-\Delta {\widetilde m}(t)m(t)\right)
    -\lim_{n \rightarrow 0}\frac{1}{n}\left(\log {\rm Tr} \exp(-H_{{\rm
stat}}) + \langle V \rangle_{{\rm stat}} \right).  \label{eq:f(t)}
\end{eqnarray}
The brackets $\langle\cdots\rangle_{{\rm stat}}$ denote the average
by the weight $\exp(-H_{{\rm stat}})$.
The equation satisfied by each order parameter can be obtained by
taking a functional derivative
with respect to the time-dependent part of the conjugate variable. The
results are
\begin{subequations}
\begin{align}
m(t)=\lim_{n \rightarrow 0}\frac{1}{n}\left\langle
   \sum_{\mu}\sigma_{t}^{\mu}\right\rangle_{{\rm stat}},
   \label{eq:state beyond SA m}  \\
q(t,t')=-\lim_{n \rightarrow 0}\frac{1}{n}\left\langle \sum_{\mu \neq
   \nu}\sigma_{t}^{\mu}\sigma_{t'}^{\nu}\right\rangle_{{\rm stat}},
   \label{eq:state beyond SA q}   \\
R(t,t')=\lim_{n \rightarrow 0}\frac{1}{n}\left\langle
   \sum_{\mu}\sigma_{t}^{\mu}\sigma_{t'}^{\mu}\right\rangle_{{\rm
stat}}. \label{eq:state beyond SA c}
\end{align}
\end{subequations}
The new Boltzmann factor $\exp(- H_{ {\rm stat} })$ is identical to
$\exp(-H_{{\rm eff}})$ under the
SA and can be simplified by the Hubbard-Stratonovich
transformation as eq. (\ref{eq:Heff}).
  From eqs. (\ref{eq:Heff}) and (\ref{eq:state beyond SA m})-(\ref{eq:state
beyond SA c}), we can verify that $q$ and
$m$ are time-independent and only $R$ is time-dependent because of
the independence of each replica in $H_{{\rm stat}}$ and the
translational invariance in the Trotter direction. Then,
   the problem is
reduced to the evaluation of the correlation function of the
one-dimensional
Ising system in a field. Calculations are somewhat involved but
straightforward. Details are given in Appendix B. The result is
\begin{equation}
R(\tau)=\int Dz_{1} \, Y^{-1}
   \left( \int Dz_{2} \, A^2\omega^{-2}\cosh\omega +
    \beta^2 \Gamma^2 \int Dz_{2} \, \omega^{-2}\cosh \omega(1-2\tau)
\right).\label{eq:c(t)}
\end{equation}
We have used the same notation as in eqs. (\ref{eq:A})-(\ref{eq:omega})
and the continuous-time notation
\begin{equation}
\tau \equiv \lim_{M \rightarrow \infty}\frac{t-t'}{M}.
\end{equation}
The result (\ref{eq:state c}) of the SA is reproduced
by integrating eq. (\ref{eq:c(t)}) over $\tau$.
The equations of other order parameters are identical to the result
under the SA. Hence, from eq. (\ref{eq:c dainari q}), we see
that the inequality $R(t,t') \geq q$ still holds for any $\tau$.

\subsection{Explicit solution of $R (\tau)$ in the ferromagnetic phase}
In the F phase,
all conjugate variables go to infinity in the limit
$p \rightarrow \infty$ and
it is reasonable to assume
$2{\widetilde R }={\widetilde q}$
according to eqs. (\ref{eq:state cc})
and (\ref{eq:state cq}). Hence, the integration over
$z_2$ just gives 1 and we find
\begin{equation}
R_{\rm F}(\tau)=\int Dz_{1}
   \left( A^2\omega^{-2} +
    \beta^2 \Gamma^2 \omega^{-2}
    \frac{\cosh \omega(1-2\tau)}{\cosh \omega} \right).
   \label{eq:cf(t)}
\end{equation}
The conjugate variables
   ${\widetilde m}$ and ${\widetilde q}$
are proportional to $p$ and very large which enables
us to derive the leading finite-$p$ correction by
the systematic large-$p$ expansion\cite{Dobr,Cesa2}.
For large conjugate parameters,
we can estimate the integral (\ref{eq:cf(t)}) by the
saddle-point method for fixed $\tau$.
To compare the time-dependent result with that under the SA,
we start from the result of the SA. Under the
condition $2{\widetilde R}={\widetilde q}$,
eq. (\ref{eq:state c}) reads
\begin{equation}
R=\int Dz_{1}
   \left( A^2\omega^{-2} +
    \beta^2 \Gamma^2 \int Dz_{2} \, \omega^{-3}
    \frac{\sinh \omega}{\cosh \omega} \right).
\label{eq:cc}
\end{equation}
The first term on
the right-hand side of eq. (\ref{eq:cc}) is rewritten as
\begin{equation}
\int Dz_{1} \,A^{2}\omega^{-2}=\int Dz_{1} \,
   \frac{\left(
	1+\sqrt{ {\widetilde q} }z_{1}/{\widetilde m}
         \right)^2
   }
{1+2 \sqrt{ {\widetilde q} } z_{1}/{\widetilde m}
   +\left(\sqrt{ {\widetilde q} }z_{1}/{\widetilde m}\right)^2
   +\left(\beta \Gamma /{\widetilde m}\right)^2.
   } \label{eq:c1}
\end{equation}
For large $p$, because $1/{\widetilde m} \propto 1/p$
is very small,
we can expand the right-hand side of eq. (\ref{eq:c1}).
After straightforward
calculations, we obtain the leading $1/p$-correction term as
\begin{equation}
   \int Dz_{1} \, A^{2}\omega^{-2}\approx 1-\frac{\beta^2
\Gamma^2}{{\widetilde m}^2}
    =1-\left(\frac{\Gamma}{j_{0}}\right)^2\frac{1}{p^2}m^{-2p+2}.
\label{eq:cmain}
\end{equation}
Similarly, the second term on
the right-hand side of eq. (\ref{eq:cc}) is rewritten as
\begin{equation}
\int Dz_{1} \,
   \left(\beta^2 \Gamma^2 \omega^{-3} -2\beta^2
\Gamma^2\omega^{-3}\frac{{\rm e}^{-2\omega}}{1+{\rm e}^{-2\omega}}\right),
   \label{eq:c2}
\end{equation}
and $\exp(-2\omega)$ is exponentially small
for large $p$. The first term of
eq. (\ref{eq:c2}) can be evaluated by the series expansion with respect to
$1/p$ as in eq. (\ref{eq:cmain}). The result is
\begin{equation}
   \int Dz_{1} \,\beta^2 \Gamma^2 \omega^{-3}\approx
     \frac{\Gamma^2}{\beta j_{0}^3} \frac{1}{p^3} m^{-3p+3}.
\end{equation}
This term is at most proportional to $1/p^3$ and is negligible. We can also
evaluate other order parameters $m$ and $q$.  Under the
condition $2{\widetilde R}={\widetilde q}$,
eq. (\ref{eq:state m}) is rewritten as
\begin{eqnarray}
m=\int Dz_{1} \, Y^{-1}\int Dz_{2} \, A\omega^{-1} \sinh \omega
   =\int Dz_{1} \, A \omega^{-1} \tanh\omega  \notag \\
   \approx \int Dz_{1} \,A\omega^{-1} \approx
    1-\frac{1}{2}\left(\frac{\Gamma}{j_{0}}\right)^2\frac{1}{p^2}m^{-2p+2}
    \approx 1-\frac{1}{2}\left(\frac{\Gamma}{j_{0}}\right)^2\frac{1}{p^2}.
\end{eqnarray}
Similarly, eq. (\ref{eq:state q}) reads
\begin{eqnarray}
q=\int Dz_{1} \, Y^{-2}
   \left( \int Dz_{2} \, A\omega^{-1} \sinh \omega \right)^2 \notag
   =\int Dz_{1} \,\left( A\omega\tanh \omega \right)^2 \\
   \approx \int Dz_{1} \, A^2 \omega^{-2} \approx
1-\left(\frac{\Gamma}{j_{0}}\right)^2\frac{1}{p^2}=R.
\end{eqnarray}
This equation is consistent with the condition
$2{\widetilde R}={\widetilde q}$.

Next, we proceed to a time-dependent analysis. From
eq. (\ref{eq:cf(t)}), its first term on the right-hand side
gives the same corrections as in the SA case.
We rewrite the second term as
\begin{equation}
\int Dz_{1} \,  \omega^{-2}
    \frac{\cosh \omega(1-2\tau)}{\cosh \omega}
   =\int Dz_{1} \, \omega^{-2}
    \frac{ {\rm e}^{-2\tau \omega} +{\rm e}^{-\omega(2-2\tau)} }{ 1+{\rm e}^{-2\omega} }
   \approx \int Dz_{1} \, \omega^{-2} \left(
    {\rm e}^{-2\tau \omega} +{\rm e}^{-\omega(2-2\tau)} \right). \label{eq:saddle}
\end{equation}
This integration can be evaluated by the saddle-point method. The
saddle-point condition of $\omega^{-2}\exp(-2\tau \omega)$ is
\begin{equation}
z^2=4\tau^2 {\widetilde q}\left(1-\frac{\beta^2 \Gamma^2}{\omega^{2}}
			  \right). \label{eq:saddle-point}
\end{equation}
It is difficult to solve this equation exactly. However, for large $p$,
the second term on the right-hand side of eq. (\ref{eq:saddle-point}) is
vanishingly small and we can approximate the saddle point as
$z=\pm 2\tau \sqrt{ {\widetilde q} }$.
The contribution from the saddle point
$z=2\tau \sqrt{ {\widetilde q}}$ is
\begin{eqnarray}
&&\!\!\!\!\!\!\!\!\!\!\!\!\!\!\!\!\!\!\!\!\!\left\{\left( 2\tau
{\widetilde q} +{\widetilde m}\right)^2
	   +\beta^2 \Gamma^2 \right\}^{-1}
    \exp\left(-2\tau^2 {\widetilde q}-2\tau
         \sqrt{\left( 2\tau {\widetilde q} +{\widetilde m}\right)^2
         +\beta^2 \Gamma^2 }\right) \notag \\
   &&\;\;\;\;\;\;  \approx \frac{1}{\beta^2 j_{0}^2}\frac{1}{p^2}
    \exp\left\{-\left(\tau^2J^2 \beta^2 +2\tau
	       \sqrt{\left(\tau^2J^2 \beta^2+\beta j_{0}\right)
	       +\frac{\beta^2 \Gamma^2}{p^2} }\right)p\right\}.
\end{eqnarray}
The other saddle
point $z=-2\tau \sqrt{ {\widetilde q} }$ gives a similar contribution.
Replacing $2\tau$ by $2-2\tau$,
we can also obtain the saddle-point
value of $\omega^{-2}\exp(-\omega(2-2\tau))$. Then,
the explicit result of $R(\tau)$ in the F phase is given by
\begin{equation}
R_{\rm F}(\tau) \approx 1-\left( \frac{\Gamma}{j_{0}} \right)^{2}\frac{1}{p^2}
+\left( \frac{\Gamma}{j_{0}} \right)^{2}\frac{1}{p^2}f(\tau,p)
\label{eq:cfex}
\end{equation}
where the function $f(\tau,p)$ expresses the time-dependent correction
which decreases exponentially as $p$ grows.
Because the third term of
eq. (\ref{eq:cfex})
is vanishingly small, the main finite-$p$ correction is
   the second term, which is identical to the SA result.
Accordingly, in the F phase,
the time-dependent part of the finite-$p$
correction
is exponentially small for large $p$
as in the CP and SG phases\cite{Dobr,Cesa2} and the SA
is valid in that sense.
We also calculated the free energy to the order $1/p$.
However, we found
that the first order correction vanishes and the
free energy remains as in
eq. (\ref{eq:fef}).
To obtain the leading order in $1/p$, we must
proceed to the next order approximation
but it is beyond our purpose in this paper.

\subsection{Remarks}
In the F phase, we have found
that the leading corrections of
order parameters are
actually time independent.
For other phases, previous works revealed that
in the CP and SG phases similar results
hold\cite{Dobr,Cesa2}.
Hence, in these instances,
the SA is valid not only in the limit
$p \rightarrow \infty$ but also
as long as $p$ is adequately large.
Meanwhile, in the QP phase, strong disagreement occurs
between the SA and the time-dependent analysis for large but finite
$p$\cite{Dobr}.
Not only the spin autocorrelation function $R(\tau)$ is actually
time dependent, but also
the low temperature behaviour shows violations of the
thermodynamic law within the SA.
  From these facts, we may conclude that the SA well
describes the region in which
quantum fluctuations are weak, but
strong quantum effects lead to a collapse
of this approximation and unphysical
behaviours of thermodynamic quantities.
In spite of such inexpediency,
the free energy recovers the SA results
in the limit $p \rightarrow \infty$ even in the QP phase.
Consequently, the SA gives correct free energies
in all the phases in the limit $p \rightarrow \infty$,
and the phase diagram depicted in
Figs. \ref{fig:p-d}-\ref{fig:T-j2-p-d} should be
exact.


\section{Conclusion}\label{sec4}
In this paper, we have studied the $p$-spin-interacting spin glass model
with ferromagnetic bias
in a transverse field by the replica method.
Trotter decomposition has been employed to reduce the quantum system
to a classical one and the SA has been assumed to obtain the
solutions of the equations of state.
We have clarified the structure of the full phase diagram, which
consists of four phases, the CP, QP, F, and SG phases.

We have also checked the validity of the SA in the F phase by the
large-$p$ expansion.
Leading finite-$p$ corrections of the order parameters have been
calculated and it has been shown that they are actually
time-independent.
It is known that similar results hold in the CP and SG phases.
This is not the case for the QP phase. Nevertheless
the free energy of the QP phase turns out to be
identical to the SA result in the limit $p \rightarrow \infty$.
In that sense, the SA gives correct solutions in this limit
and our phase diagram is exact.

Admittedly, the model investigated in this paper
is not a faithful reproduction of real SG systems.
However, it has a great advantage that order parameters and the free
energy can be exactly obtained.
Goldschmidt found
two qualitatively different types of paramagnetic phases, CP and QP.
We also found that quantum fluctuations reduce the ferromagnetic order
and cause a transition to the QP phase.
These properties appear to be plausible in more realistic systems.
On the other hand,
we saw that in the CP, F, and SG phases
quantum fluctuations are completely irrelevant,
which should be specific to this model.
For more realistic models (like the SK model) we should take into
account the influence of quantum fluctuations on order parameters.
However, it is difficult to treat such an effect as was done in the
present paper and we need different techniques.
Finding effective approaches and improving the SA remain interesting
problems to be investigated in the future. Our present results will
serve as a first step to understanding the interplay between quantum
fluctuations, ferromagnetic bias and quenched disorder.

After submission of the manuscript, we came to notice that the same
problem was discussed by Saakyan\cite{Saak} and Jun-Ichi
Inoue\cite{QUAN}.
The originality of our work
lies in the systematic analysis of the validity of the SA and the
explicit clarification of the structure of the full phase diagram.


\acknowledgement
   It is a pleasure to thank Dr. Kazutaka Takahashi for useful discussions and
   suggestive comments on the manuscript. We also indebt Prof. D. Saakyan
   for letting us know his paper.
This work was supported by the Grand-in-Aid for Scientific Research on
the Priority Area ``Deepening and Expansion of Statistical Mechanical
Informatics'' by  the Ministry of Education, Culture, Sports, Science
and Technology as well as by the CREST, JST.


\appendix

\section{1RSB free energy and the inequalities $R \geq q_{1} \geq q_{0}$ }

In this Appendix we derive the 1RSB free energy eq. (\ref{eq:fe1rsb})
and the inequalities $R \geq q_{1} \geq q_{0}$.
Under the 1RSB scheme and the SA,
eq. (\ref{eq:rep fe}) reads
\begin{eqnarray}
\beta f = \frac{1}{2}m_{1}
   \left(\frac{1}{2}\beta^{2}J^{2}q_{0}^p-{\widetilde q_{0}}q_{0} \right)
    -\frac{1}{2}(m_{1}-1)
   \left(\frac{1}{2}\beta^{2}J^{2}q_{1}^p-{\widetilde q_{1}}q_{1} \right)
   -
   \left(\frac{1}{4}\beta^{2}J^{2}R^p-{\widetilde R}R \right) \notag\\
   -
   \left(\beta j_{0}m^p-{\widetilde m}m \right)
   -\lim_{M \rightarrow \infty}\lim_{n \rightarrow 0}\frac{1}{n} \log
{\rm Tr} \exp\left(-H_{{\rm eff}}\right).
\end{eqnarray}
The effective Hamiltonian $H_{{\rm eff}}$ is written in the form
\begin{eqnarray}
   &&\!\!\!\!\!\!\!\!\!\!\!\!\!\!\!\!\!\!\!\!\!
    -H_{{\rm eff}}= B\sum_{\mu}\sum_{t}\sigma^{\mu}_{t}\sigma_{t+1}^{\mu}
    +{\widetilde m}
    \sum_{\mu}\sum_{t}\sigma^{\mu}_{t}
    +{\widetilde R}
    \sum_{\mu}\sum_{t \neq t'}\sigma^{\mu}_{t}\sigma_{t'}^{\mu} \notag \\
   &&\; \; \; \; +\frac{1}{2}\left(
			    {\widetilde q_{0}}
			    \sum_{\mu \neq \nu}\sum_{t,t'}
			    \sigma^{\mu}_{t}\sigma_{t'}^{\nu}
			    +\left({\widetilde q_{1}}-{\widetilde q_{0}}\right)
			    \sum_{l}^{ n/m_{1} }\sum_{\mu_{l} \neq \nu_{l}}^{\rm block}
			    \sum_{t,t'}
			    \sigma^{\mu_{l}}_{t}\sigma_{t'}^{\nu_{l}}
			   \right).\label{eq:Heffrsb}
\end{eqnarray}
We can rewrite the summation of spin products as
\begin{eqnarray}
   &&-H_{{\rm eff}}=
\frac{1}{2}{\widetilde q_{0}}\left(\sum_{\mu}\sum_{t} \sigma_{t}^{\mu}\right)^2
+\frac{1}{2}\left({\widetilde q_{1}}-{\widetilde q_{0}}\right)\sum_{l}^{n/m_{1}}
\left\{ \left(\sum_{\mu_{l}}\sum_{t}\sigma_{t}^{\mu_{l}}\right)^2-
   \sum_{\mu_{l}}\left(\sum_{t}\sigma_{t}^{\mu_{l}}\right)^2
\right\} \notag
\\
&&+\left({\widetilde R}-\frac{1}{2}{\widetilde
q_{0}}\right)\sum_{\mu}\left(\sum_{t} \sigma_{t}^{\mu}\right)^2
+B\sum_{\mu}\sum_{t}\sigma_{t}^{\mu}\sigma_{t+1}^{\mu}+{\widetilde
m}\sum_{\mu}\sum_{t}\sigma_{t}^{\mu}. \label{eq:Heffrsb2}
\end{eqnarray}
To take the spin trace, the Hubbard-Stratonovich
transformation is employed for the quadratic terms. The result is
\begin{eqnarray}
   {\rm e}^{-H_{{\rm eff}}}=  \int Dz_{1} \, \prod_{l}^{n/m_{1}}\left\{
\int Dz_{2}
\prod_{\mu_{l}=(l-1)m_{1}+1}^{lm_{1}}
\left( \int Dz_{3} \, {\rm e}^{L} \right) \right\}, \label{eq:rsbhs}
\end{eqnarray}
where
$L \equiv
A_{1}/M\sum_{t}\sigma_{t}^{\mu_{l}}+B\sum_{t}\sigma_{t}^{\mu_{l}}\sigma_{t+1}^{\mu_{l}}$
and $A_{1}$ is defined in eq. (\ref{eq:A1}).
Using the Trotter formula, we can take the limit
$M \rightarrow \infty$ and perform the spin trace as in
eq. (\ref{eq:Trotter}).
The result in the limit $n \rightarrow 0$ is
given as
\begin{eqnarray}
\lim_{M \rightarrow \infty}\lim_{n \rightarrow 0}\frac{1}{n} \log {\rm Tr}
\exp\left(-H_{{\rm eff}}\right) =
\frac{1}{m_{1}}\int Dz_{1} \,\log \int Dz_{2}\left(\int Dz_{3} \,
2\cosh \omega_{1} \right)^{m_{1}}.
\end{eqnarray}
The equations of state
eqs. (\ref{eq:state cmrsb})-(\ref{eq:state q1rsb}) are
obtained by taking a variation of the free energy
eq. (\ref{eq:fe1rsb})
with respect to conjugate variables and order parameters.

Next, we derive the inequalities $R \geq q_{1} \geq q_{0}$.
   From eqs. (\ref{eq:state mrsb})-(\ref{eq:state q1rsb}),
\begin{eqnarray}
R&=&\int Dz_{1} \, Y^{-1}_{1}
   \int Dz_{2} \, Y_{2}^{m_{1}-1}\left(\int Dz_{3} \,
		   A^2_{1}\omega^{-2}_{1} \cosh\omega_{1} +
    \beta^2 \Gamma^2 \int Dz_{3} \, \omega^{-3}_{1}\sinh \omega_{1} \right) \notag \\
&&\geq \int Dz_{1} \, Y^{-1}_{1}
   \int Dz_{2} \, Y_{2}^{m_{1}-1} \int Dz_{3} \,
		   A^2_{1}\omega^{-2}_{1} \cosh\omega_{1}  \notag \\
&& \geq \int Dz_{1} \, Y^{-1}_{1}
   \int Dz_{2} \, Y_{2}^{m_{1}-1} \int Dz_{3} \,
		   A^2_{1}\omega^{-2}_{1} \sinh \omega_{1} \notag \\
&&\geq \int Dz_{1} \,Y^{-1}_{1}
   \int Dz_{2} \, Y_{2}^{m_{1}-2} \left(\int Dz_{3} \,
		   A_{1}\omega^{-1}_{1} \sinh \omega_{1} \right)^2=q_{1}.
\label{eq:c(t) dainari q1}
\end{eqnarray}
Similarly, we can show $q_{1} \geq q_{0}$
from the definition of $Y_{1}$ (\ref{eq:Y1}) as
\begin{eqnarray}
q_{1}&=&\int Dz_{1} \, Y^{-1}_{1}
   \int Dz_{2} \, Y_{2}^{m_{1}-2} \left( \int Dz_{3} \,
		   A_{1}\omega^{-1}_{1} \sinh \omega_{1} \right)^2 \notag \\
&\geq&
   \int Dz_{1}\left( Y^{-1}_{1}
   \int Dz_{2} \, Y_{2}^{m_{1}-1}\int Dz_{3} \,
		   A_{1}\omega^{-1}_{1} \sinh \omega_{1} \right)^2 = q_{0}.
\end{eqnarray}

\section{Evaluations of the correlation function}
We calculate the correlation function and derive the
expression for $R(\tau)$ eq. (\ref{eq:c(t)}).
   The un-normalized correlation function of the one-dimensional Ising
   system with periodic boundary is
\begin{equation}
G(t,t')={\rm Tr} \sigma_{t}\sigma_{t'}
   \exp
\left(J\sum_{t=1}^{M}\sigma_{t}\sigma_{t+1}+h\sum_{t=1}^{M}\sigma_{t}\right).
\end{equation}
We can compute the correlation function $G(t,t')$
by the transfer matrix method.
The general solution is
\begin{equation}
G(t,t')=4x_{+}^{2}x_{-}^{2}
   \left(\lambda_{+}^{t-t'}\lambda_{-}^{M-(t-t')}
    +\lambda_{+}^{M-(t-t')}\lambda_{-}^{t-t'}
   \right)
   +(2x_{+}^2-1)^2\lambda_{+}^{M}+(2x_{-}^2-1)^2\lambda_{-}^{M},
\end{equation}
where $\lambda_{\pm}$ are the eigenvalues of the transfer matrix and
$x_{\pm}$ are the first components of the eigenvectors $| {\pm} \rangle$.
Their explicit
forms are
\begin{eqnarray}
   \lambda_{\pm}={\rm e}^{J}\left(\cosh h \pm \sqrt{\cosh^{2}h-1+{\rm e}^{-4J} }\right), \\
   | {\pm} \rangle =D_{\pm} \left(
		    \begin{array}{cc}
		     -{\rm e}^{-J} \\
		     {\rm e}^{J}\left(\sinh h \mp \sqrt{\sinh^2 h +{\rm e}^{-4J}}
			  \right)
		    \end{array}
			  \right)
   \equiv \left(
		    \begin{array}{cc}
		     x_{\pm} \\
		     y_{\pm}
		    \end{array}
			  \right),
\end{eqnarray}
where $D_{\pm}$ are normalization constants.
In the present case, the parameters
$J=B=\log \left(\coth \beta \Gamma/M \right)^{1/2}$ and $h=A/M$ depend
on the number of
spins $M$. Restoring the omitted overall factor
$C=\left\{ \left(1/2\right)\sinh 2\beta \Gamma / M \right\}^{1/2}$,
we can obtain the finite value of the correlation function in
the limit $M \rightarrow \infty$. After straightforward
calculations, we get
\begin{equation}
x_{\pm}^2 \rightarrow \frac{1}{2}\frac{\left(\beta
\Gamma\right)^2}{\omega^2 \mp A\omega} \ , \
(C\lambda_{\pm})^{M} \rightarrow {\rm e}^{\pm \omega},
\end{equation}
where $\omega$ is given in eq. (\ref{eq:omega}).
Hence, the correlation function $G(t,t')$ is given by
\begin{equation}
G(t,t') \rightarrow
G(\tau) =2A^2\omega^{-2}\cosh \omega +2(\beta \Gamma)^2
\omega^{-2}\cosh \omega(1-2\tau)
		   . \label{eq:cor}
\end{equation}
Substituting eq. (\ref{eq:cor}) into eq. (\ref{eq:state beyond SA c}) and
taking the
limit $n \rightarrow 0$, we finally get
\begin{equation}
R(\tau)=\int Dz_{1} \, Y^{-1}
   \left( \int Dz_{2} \, A^2\omega^{-2}\cosh\omega +
    \beta^2 \Gamma^2 \int Dz_{2} \, \omega^{-2}\cosh \omega(1-2\tau) \right) \notag.
\end{equation}

\end{document}